\documentclass[%
preprint%
 ,secnumarabic%
,amssymb, amsmath,nobibnotes,aps]{revtex4}

\usepackage{epsfig}%
\usepackage{graphicx}%
\expandafter\ifx\csname package@font\endcsname\relax\else
 \expandafter\expandafter
 \expandafter\usepackage
 \expandafter\expandafter
 \expandafter{\csname package@font\endcsname}%
\fi

\begin{document}

\title{Non-Gaussianity as a signature of thermal initial condition of inflation}%

\author{Suratna Das and Subhendra Mohanty}%
\affiliation{Physical Research Laboratory, Ahmedabad 380009,
India.}
\def\be{\begin{equation}}
\def\ee{\end{equation}}
\def\al{\alpha}
\def\bea{\begin{eqnarray}}
\def\eea{\end{eqnarray}}

\begin{abstract}
We study non-Gaussianities in the primordial perturbations in single
field inflation where there is radiation era prior to
inflation. Inflation takes place when the energy density of radiation
drops below the value of the potential of a coherent scalar field.  We
compute the thermal average of the two, three and four point
correlation functions of inflaton fluctuations.  The three point
function is proportional to the slow roll parameters and there is an
amplification in $f_{NL}$ by a factor of $65$ to $90$ due to the
contribution of the thermal bath, and we conclude that the bispectrum
is in the range of detectability with the 21-cm anisotropy
measurements. The four point function on the other hand appears in
this case due to the thermal averaging and the fact that thermal
averaging of four-point correlation is not the same as the square of
the thermal averaging of the two-point function. Due to this fact
$\tau_{NL}$ is not proportional to the slow-roll parameters and can be
as large as $-42$. The non-Gaussianities in the four point correlation
of the order $10$ can also be detected by 21-cm background
observations. We conclude that a signature of thermal inflatons is a
large trispectrum non-Gaussianity compared to the bispectrum
non-Gaussianity.
\end{abstract}
\maketitle
\section{Introduction}
The experimental determination of signature of primordial
non-Gaussianity in the CMB spectrum is of great interest as
non-Gaussianity can give us insights into the dynamics of inflation
models \cite{matarrese, Komatsu:2009}. Interacting fields have
non-Gaussian correlations proportional to the coupling strength, which
for the inflaton is unmeasurably small \cite{raghu}. In quasi
de-Sitter space, single field inflation models predict a level of
non-Gaussianity proportional to the slow roll parameters
\cite{maldacena, Acquaviva, Chen}. The current constraint on CMB
bispectrum from the WMAP 5yr data is $-151 < f_{NL}^{eq}<253$ ($95 \%$
CL) \cite{Komatsu}, $f_{NL}$ being the bispectrum non-linear parameter
and a measure of non-Gaussianity. The single field inflation models
prediction of $f_{NL} \sim 10^{-2}$ is too small to be detectable in
WMAP or the upcoming PLANCK mission where non-Gaussianities at the
level of $f_{NL} \sim 5$ can be probed \cite{spergel}. The most
sensitive probe of primordial non-Gaussianities can come from the
measurement of anisotropies of the 21-cm background whose bispectrum
can probe $f_{NL}< 0.1$ \cite{cooray,cooray-prd}. WMAP constrains the
non-Gaussianity from trispectrum at
$\left|\tau_{NL}\right|\lesssim10^8$ \cite{WMAP-tri} while PLANCK is
expected to reach the sensitivity upto $\left|\tau_{NL}\right|\sim
560$ \cite{PLANCK-tri}. The anisotropies of the 21-cm background can
constrain the trispectrum of primordial perturbations to the level of
$ \sim 10$ \cite{cooray-prd}, which is still too large compared to the
predictions of the single field models of $\tau_{NL} \simeq (\frac65
f_{NL})^2$ \cite{seery, wands}.

It was shown by Gangui et al. \cite{gangui} and more recently in
\cite{holman, Chen} that if the initial state of the inflatons is not
the Bunch-Davies vacuum but some excited state then there is an
enhancement of the non-Gaussianity from such initial state effects.  A
natural example of a non-Bunch Davies initial state arises if there is
a pre-inflation radiation era prior to inflation \cite{prl1}.
Inflation takes place when the energy density of radiation $\rho_r$
drops below the value of the potential of a coherent scalar field. In
such models it is seen that the power-spectrum is enhanced at low $k$
which can be used to put constraint on the comoving temperature at the
time of inflation \cite{prl1}. Inflation scenario with a pre-radiation
era have an interesting prediction that the B-mode polarization
spectrum is enhanced at low $l$ due the contribution of thermal
gravitons \cite{prl2,coles}.

The scenario of thermal initial condition is very general and would be
applicable for any model of inflation if there was a pre-inflationary
radiation dominated era. The effects of the initial thermal era to be
observable either in the CMB anisotropy spectrum or in the
non-gaussianities the perturbations entering the horizon today should
have left the de-Sitter horizon at a temperature $T$ not too small
compared to $H$ (the Hubble parameter at the time of inflation). If
there were a large number of e-foldings prior to the present
perturbation modes leaving the inflation horizon then the effect of
the pre-inflationary thermal era would be unobservable. In models
where the total number of e-foldings are just enough to solve the
flatness and horizon problems, there can be a imprint of the spatial
curvature at the time of inflation on the power spectrum. This has
been studied in \cite{masso}. A natural model where inflation
commences just as the temperature falls below a critical temperature
and is of limited duration is where a fermion pair forms a scalar
condensate which acts as the inflaton. Such models have been studied
in \cite{das, right}.

In this paper we study non-Gaussianities in the primordial
perturbation in single field inflation where there is radiation era
prior to inflation. The thermal background of inflaton, gravitons and
other fields is decoupled from the actual dynamical evolution of the
inflaton unlike in the warm inflation models \cite{warm}, where there
can be large non-Gaussianities \cite{moss,Chen-thermal} due to
dissipative coupling between the inflaton and the radiation bath. In
this model the temperature of the decoupled radiation bath goes down as
$T_{\rm ph}=T/a$ where $T$ is the constant comoving temperature. The
thermal distribution functions which depend on the ratio $\frac{k_{\rm
    ph}}{T_{\rm ph}}=\frac kT$ (where $k$ is the comoving wavenumber
of the perturbations) retain the same form during inflation.

We calculate the thermal average of the three point correlation
function, otherwise known as the bispectrum, of the comoving density
perturbations. The non-linear parameter $f_{NL}$ being a measure of
non-Gaussianity due to bispectrum is turned out to be a function of
the magnitude of the three momenta of comoving perturbations. In order
to quantify $f_{NL}$ three distinct configurations of those momenta
are analyzed namely the ``squeezed'' triangle, the ``equilateral''
triangle and the ``folded'' triangle and it is observed that the
maximum contribution for $f_{NL}$ comes from the ``equilateral''
configuration. We find that the thermal contribution can result in the
enhancement of $f_{NL}$ by factors ranging from $65-90$.

We show that due to the presence of the initial temperature the
contribution to four point correlation function of the density
perturbations comes from different factors other than the mere
disconnected diagrams. We evaluate the thermal average of the four
point function and calculate the contribution of the thermal initial
states to $\tau_{NL}$. We find that in the leading order $\tau_{NL}$
is independent of the slow roll parameters and we find that it can be
as large as $ \tau_{NL} \sim -42$.

We organize the rest of the paper as follows. In Section~\ref{sec-2},
we compute the thermal average of the two point correlation of the
density perturbations. We see that the two point function is enhanced
at low $k$ due to the contribution of the thermal inflatons as
observed in \cite{prl1}. In Section~\ref{sec-3}, we compute the
thermal averaged three-point correlations and calculate $f_{NL}$ for
various momentum configurations. In Section \ref{sec-4} we compute the
thermal contribution to $\tau_{NL}$. We discuss the feasibility of
measuring the non-Gaussianity due to thermal inflatons in the
Concluding section. We have given a outline of the calculation of the
non-Gaussianity parameter $f_{NL}$ at zero temperatures for the single
field slow-roll inflation models in the Appendix.

\section{Thermal average of inflaton power spectra}
\label{sec-2}

If there was a radiation era prior to inflation one expects a thermal
distribution of inflatons to be present which might have decoupled
from other fields prior to inflation. It has been shown in \cite{prl1}
that this thermal distribution of inflaton modifies the power spectrum
of inflaton fluctuations and the curvature power spectrum will have an
additional temperature depended term. In this section we compute the
two point correlation of inflaton perturbations taking this thermal
distribution of inflatons into consideration.

The Fourier expansion of inflaton fluctuations in de-Sitter space is
\begin{eqnarray}
\delta\phi({\mathbf x},t)=\int\frac{d^3{\mathbf k}}{\left(2\pi\right)^{\frac32}}\left(b_{\mathbf k}\varphi_k(t)+b^{\dagger}_{-\mathbf k}\varphi^{\ast}_k(t)\right)e^{i{\mathbf k}\cdot{\mathbf x}},
\end{eqnarray}
where $\varphi_k(t)$ are the mode functions which satisfy the
Klein-Gordon equation in Fourier space and $b_{\mathbf k}$ and
$b^{\dagger}_{\mathbf k}$ are the annihilation and creation operators
respectively. In Fourier space the inflaton fluctuations can be
written as
\begin{eqnarray}
\delta\phi({\mathbf k},t)=b_{\mathbf k}\varphi_k(t)+b^{\dagger}_{-\mathbf k}\varphi^{\ast}_k(t).
\label{phik}
\end{eqnarray}
The canonical commutation relation satisfied by these creation and
annihilation operators is
\begin{eqnarray}
\left[b_{\mathbf k_1},b^{\dagger}_{\mathbf k_2}\right]=\delta^3(\mathbf{k_1}-{\mathbf k_2}),
\label{commutation}
\end{eqnarray}
with the vacuum satisfying $b_{\mathbf k}|0\rangle=0$ at zero
temperature, which ensures that the vacuum has zero occupation
$N_k|0\rangle=0$ where $N_k\equiv b^{\dagger}_{\mathbf k}b_{\mathbf
  k}$ is the number operator. The power spectrum of inflaton
$P_{\delta\phi}(k)$ is the two-point correlation function of the
inflaton fluctuations in momentum space which is defined as
\begin{eqnarray}
P_{\delta\phi}(k)\equiv\frac{k^3}{2\pi^2}\left\langle\delta\phi(k,t)\delta\phi(k,t)\right\rangle,
\end{eqnarray}
where $k\equiv|{\mathbf k}|$.

In this case, where there is a radiation era prior to inflation, the
inflaton will have a thermal distribution during inflation. Due to
this distribution the thermal vacuum $|\Omega\rangle\equiv
|n_{k_1},n_{k_2},\cdots\rangle$ will now contain real particles
yielding
\begin{eqnarray}
N_k|\Omega\rangle=n_k|\Omega\rangle,
\label{thermal-vacuum}
\end{eqnarray}
where $n_k$ is the number of particles with momentum ${\mathbf k}$
present in the thermal vacuum. In general, for creation-annihilation
operators with different momenta one gets
\begin{eqnarray}
b^{\dagger}_{\mathbf k_1}b_{\mathbf k_2}|\Omega\rangle=\delta^3({\mathbf k}_1-{\mathbf k}_2)n_{k_1}|\Omega\rangle.
\end{eqnarray}
Throughout this paper we will consider non-interacting real scalar
field for which the chemical potential $\mu=0$. For a single inflaton
with momentum ${\mathbf k}$ the partition function will be
\begin{eqnarray}
z=\sum_{n_{k}=0}^{\infty}e^{-\beta n_{k}k}=\frac{1}{1-e^{-\beta k}}\, ,
\label{z}
\end{eqnarray}
where $\beta$ is the inverse of the comoving  temperature $T$.
 Due to this thermal distribution of the inflaton fluctuation 
 a thermal statistical average of the two-point correlation
function will determine the power spectrum 
\begin{eqnarray}
P^{\rm th}_{\delta\phi}(k)&=&\frac{k^3}{2\pi^2}\left\langle\Omega|\delta\phi(k,t)\delta\phi(k,t)|\Omega\right\rangle_\beta \nonumber \\
&=&\frac{k^3}{2\pi^2}\sum_{\varepsilon_k}{p}(\varepsilon_k)\left\langle\Omega|\delta\phi(k,t)\delta\phi(k,t)|\Omega\right\rangle.
\label{p-th}
\end{eqnarray}
Here ${p}(\varepsilon_k)$ is the probability of the system to be in
the state $\varepsilon_k\equiv n_k k$ which is defined as
\begin{eqnarray}
{p}(\varepsilon_k)\equiv\frac{e^{-\beta n_k k}}{\sum_{n_k}e^{-\beta
n_k k}}=\frac{e^{-\beta n_k k}}{z}\, ,
\end{eqnarray}
where $z$ is given in Eq.~(\ref{z}). However, due to the thermal
distribution of the inflaton field the inflaton fluctuations will
follow the relations given in Eq.~(\ref{commutation}) and
Eq.~(\ref{thermal-vacuum}) which yield
\begin{eqnarray}
\langle\Omega|\delta\phi(k,t)\delta\phi(k,t)|\Omega\rangle&=&\left|\varphi_{k}(t)\right|^2\left\langle\Omega|\left(1+2N_{k}\right)|\Omega\right\rangle\nonumber\\
&=&\left|\varphi_{k}(t)\right|^2\left(1+2n_{k}\right).
\end{eqnarray}
Hence the power spectrum given in Eq.~(\ref{p-th}) will be
\begin{eqnarray}
P^{\rm th}_{\delta\phi}(k)&=&\frac{k^3}{2\pi^2}\left|\varphi_{k}(t)\right|^2\frac{1}{z}\sum_{n_k}e^{-\beta n_k k}\left(1+2n_{k}\right)\nonumber \\
&=&\frac{k^3}{2\pi^2}\left|\varphi_{k}(t)\right|^2(1+2f_B(k)),
\label{p-th-1}
\end{eqnarray}
where $f_B(k)\equiv\frac{1}{e^{\beta k}-1}$ is the Bose-Einstein
distribution. To get the last equality in the above equation the
following relation is used \cite{gradstein}
\begin{eqnarray}
\sum_{n=0}^{\infty}nq^n=\frac{q}{\left(1-q\right)^2}.
\label{identity}
\end{eqnarray}
Now for a light inflaton ($m_{\phi}\ll H$, $m_{\phi}$ being the mass
of the inflaton and $H$ being the Hubble parameter during inflation)
the mode function has the solution \cite{Riotto_lec}
\begin{eqnarray}
\left|\varphi_{k}\right|\simeq\frac{H}{\sqrt{2k^3}}\left(\frac{k}{aH}\right)^{\frac32-\nu_\varphi},
\end{eqnarray}
where $a$ is the cosmic scale factor and
$\nu_{\varphi}\simeq\frac32-\frac{m_{\phi}^2}{H^2}$. In a generic
single field inflationary model this mode function solution along with
the $k^3$ factor in the power spectrum gives a nearly scale invariant
spectra for inflaton fluctuations. But due to the thermal distribution
of the inflaton fluctuations, expression for power spectrum in
Eq.~(\ref{p-th-1}) contains an additional temperature dependent
factor of $(1+2f_B(k))=\coth(\beta k/2)$. Thus the thermal power
spectrum of inflaton fluctuations is given by
\begin{eqnarray}
P^{\rm th}_{\delta\phi}(k)=\coth(\beta k/2)P_{\delta\phi}(k),
\end{eqnarray}
and hence the thermal average of the power spectrum for comoving
curvature perturbations defined in Eq.~(\ref{curvature}) will be
\begin{eqnarray}
{\mathcal P}^{\rm th}_{\mathcal R}(k)=\coth(\beta k/2){\mathcal P}_{\mathcal R}(k),
\end{eqnarray}
as has been already stated in \cite{prl1}.

In \cite{prl1} the CMB power spectrum generated using the thermal
comoving curvature power spectrum is compared with WMAP data and a
constraint on comoving temperature is given as $T<1.0\times 10^{-3}$
Mpc$^{-1}$ (with the convention that the present scale factor $a_0
\equiv 1$).  Such a bound is also found in \cite{coles} from thermal
primordial gravitational waves. Since $T=a_i T_{\rm ph}$ where $T_{\rm
  ph}$ and $a_i$ are the physical temperature and the scale factor
when our current horizon scale crossed the de-Sitter horizon during
inflation, this constraint can be rewritten as $T_0<4.2H$. As the
comoving wavenumber $k=a_i H$ one can put a lower bound on $\beta k$
as
\begin{eqnarray}
\beta k=\frac{a_i H}{a_i T_{\rm ph} }>0.238.
\label{lower}
\end{eqnarray}
This lower bound on $\beta k$ will be used in following sections to
quantify the maximum value of non-Gaussianity in thermal bispectrum
and thermal trispectrum.
\section{Non-Gaussianity in bispectrum from thermal distribution of inflaton}
\label{sec-3}

The three point correlation function of comoving curvature
perturbations ${\mathcal R}$ or the bispectrum is defined in
Eq.~(\ref{bispec}) and the non-linear parameter for bispectrum in the
case of single field slow-roll model is given in
Eq.~(\ref{fnl-slow}). In presence of a pre-inflationary radiation era
the bispectrum will also receive a modification as in the case of the
thermal power spectrum. Hence in this case the three point correlation
function of the non-linear curvature perturbation will be
\begin{eqnarray}
\left\langle{\mathcal R}_{NL}({\mathbf k}_1){\mathcal R}_{NL}({\mathbf k}_2){\mathcal R}_{NL}({\mathbf k}_3)\right\rangle_{\beta}&=&\frac12\left(\frac{H}{\dot{\phi}}\right)^2\frac{\partial}{\partial\phi}\left(\frac{H}{\dot{\phi}}\right)\int\frac{d^3{\mathbf p}}{\left(2\pi\right)^{\frac32}}\times\nonumber \\
&&\left[\left\langle\delta\phi_L({\mathbf p})\delta\phi_L({\mathbf k}_1-{\mathbf p})\delta\phi_L({\mathbf k}_2)\delta\phi_L({\mathbf k}_3)\right\rangle_{\beta}+2 \,\,{\rm perms}\right],
\label{therm-R}
\end{eqnarray}
where R.H.S. of the above equation contains the thermal average of
four-point correlation functions of the inflaton perturbations.

We will first generalise the case of thermal average of the two-point
correlation function to derive the thermal average of the four-point
correlation function of scalar perturbations with any four
momenta. The thermal average of higher order correlation functions is
of the form
\begin{equation}
\langle \phi_{k_1} \phi_{k_2}\phi_{k_3}\cdots \rangle_\beta=
\sum_{\lbrace n_{k_i} \rbrace} { p}(k_1,k_2, k_3,\cdots)\,\langle
\Omega| \phi_{k_1} \phi_{k_2}\phi_{k_3}\cdots| \Omega \rangle,
\end{equation}
where the thermal probability of the occupancy of different momenta
${\mathbf k}_i$ and $\varepsilon\equiv \sum_{n_{k_r}}n_{k_r}k_r$ is
\begin{eqnarray}
{ p}(k_1,k_2, k_3,\cdots)\equiv\frac{\prod_re^{-\beta n_{k_r}
k_r}}{\prod_r\sum_{n_k}e^{-\beta n_{k_r}
k_r}}=\frac{\prod_re^{-\beta n_{k_r} k_r}}{Z}.
\end{eqnarray}
Here $Z$ is the grand partition function of massless inflatons with
energies $E_{k_r}=\sqrt{{\mathbf k_r}^2}=k_r$ which is given as
\begin{eqnarray}
Z=\prod_r\sum_{n_{k_r}=0}^{\infty}e^{-\beta
n_{k_r}k_r}=\prod_r\frac{1}{1-e^{-\beta k_r}}, \label{Z}
\end{eqnarray}
where $r$ is the index for different energy levels.

The four-point correlation function of inflaton fluctuations with four
different momenta contains six different combinations of two creation
and two annihilation operators and thermal average of one of these
combinations can be derived as follows :

Let us consider the thermal average of $\left(b^{\dagger}_{-\mathbf
  k_1}b_{\mathbf k_2}b^{\dagger}_{-\mathbf k_3}b_{\mathbf k_4}\right)$
which yields
\begin{eqnarray}
\left\langle b^{\dagger}_{-\mathbf k_1}b_{\mathbf k_2}b^{\dagger}_{-\mathbf k_3}b_{\mathbf k_4}\right\rangle_\beta&=&\sum_{\varepsilon}p(k_1,k_2,k_3,k_4)\left\langle\Omega\left| b^{\dagger}_{-\mathbf k_1}b_{\mathbf k_2}b^{\dagger}_{-\mathbf k_3}b_{\mathbf k_4}\right|\Omega\right\rangle\nonumber\\
&=&\delta^3({\mathbf k_1}+{\mathbf k_2})\delta^3({\mathbf k_3}+{\mathbf k_4})\frac{1}{Z}\sum_{n_{k_1}}\sum_{n_{k_2}}e^{-\beta\left(n_{k_1}k_1+n_{k_3}k_3\right)}\left[n_{k_1}n_{k_3}\right],
\end{eqnarray}
where $Z=\left(\frac{1}{1-e^{-\beta k_1}}\right)\left(\frac{1}{1-e^{-\beta k_3}}\right)$. The summations in the above equation yields
\begin{eqnarray}
\left\langle b^{\dagger}_{-\mathbf k_1}b_{\mathbf k_2}b^{\dagger}_{-\mathbf k_3}b_{\mathbf k_4}\right\rangle_\beta&=&\delta^3({\mathbf k_1}+{\mathbf k_2})\delta^3({\mathbf k_3}+{\mathbf k_4})\left[f_B(k_1)f_B(k_3)\right],
\end{eqnarray}
where the identity stated in Eq.~(\ref{identity}) is used. Similarly
the thermal average of other combinations of the two creation and two
annihilation operators will be
\begin{eqnarray}
\left\langle b_{\mathbf k_1}b_{\mathbf k_2}b^{\dagger}_{-\mathbf k_3}b^{\dagger}_{-\mathbf k_4}\right\rangle_\beta&=&\delta^3({\mathbf k_1}+{\mathbf k_4})\delta^3({\mathbf k_2}+{\mathbf k_3})\left[1+f_B(k_1)\right]+\nonumber \\
&&\delta^3({\mathbf k_1}+{\mathbf k_3})\delta^3({\mathbf k_2}+{\mathbf k_4})\left[1+f_B(k_1)+f_B(k_2)+f_B(k_1)f_B(k_2)\right], \\
\left\langle b_{\mathbf k_1}b^{\dagger}_{-\mathbf k_2}b_{\mathbf k_3}b^{\dagger}_{-\mathbf k_4}\right\rangle_\beta&=&\delta^3({\mathbf k_1}+{\mathbf k_2})\delta^3({\mathbf k_3}+{\mathbf k_4})\left[1+f_B(k_1)+f_B(k_3)+f_B(k_1)f_B(k_3)\right],\\
\left\langle b_{\mathbf k_1}b^{\dagger}_{-\mathbf k_2}b^{\dagger}_{-\mathbf k_3}b_{\mathbf k_4}\right\rangle_\beta&=&\delta^3({\mathbf k_1}+{\mathbf k_2})\delta^3({\mathbf k_3}+{\mathbf k_4})\left[f_B(k_3)+f_B(k_1)f_B(k_3)\right], \\
\left\langle b^{\dagger}_{-\mathbf k_1}b_{\mathbf k_2}b_{\mathbf k_3}b^{\dagger}_{-\mathbf k_4}\right\rangle_\beta&=&\delta^3({\mathbf k_1}+{\mathbf k_2})\delta^3({\mathbf k_3}+{\mathbf k_4})\left[f_B(k_1)+f_B(k_1)f_B(k_3)\right], \\
\left\langle b^{\dagger}_{-\mathbf k_1}b^{\dagger}_{-\mathbf k_2}b_{\mathbf k_3}b_{\mathbf k_4}\right\rangle_\beta&=&-\delta^3({\mathbf k_1}+{\mathbf k_4})\delta^3({\mathbf k_2}+{\mathbf k_3})f_B(k_1)\nonumber\\
&&+\delta^3({\mathbf k_1}+{\mathbf k_3})\delta^3({\mathbf k_2}+{\mathbf k_4})\left[f_B(k_1)f_B(k_2)\right].
\end{eqnarray}
Hence the thermal average of a general four-point correlation function
with four different momenta will be
\begin{eqnarray}
&&\left\langle\delta\phi({\mathbf k_1},t)\delta\phi({\mathbf
  k_2},t)\delta\phi({\mathbf k_3},t)\delta\phi({\mathbf k_4},t)\right\rangle_\beta=\left|\varphi_{k_1}(t)\right|^2\left|\varphi_{k_2}(t)\right|^2\left[\delta^3({\mathbf k_1}+{\mathbf k_4})\delta^3({\mathbf k_2}+{\mathbf k_3})\right.\nonumber \\
&&+\delta^3({\mathbf k_1}+{\mathbf k_3})\delta^3({\mathbf k_2}+{\mathbf k_4})
\left.\left\lbrace1+f_B(k_1)+f_B(k_2)+2f_B(k_1)f_B(k_2)\right\rbrace\right]+\left|\varphi_{k_1}(t)\right|^2\left|\varphi_{k_3}(t)\right|^2\nonumber\\
&&\times\left[\delta^3({\mathbf k_1}+{\mathbf k_2})\delta^3({\mathbf k_3}+{\mathbf k_4})\left\lbrace1+2f_B(k_1)+2f_B(k_3)+4f_B(k_1)f_B(k_3)\right\rbrace\right].
\label{fpc}
\end{eqnarray}
With this general result one can calculate the thermal average of the
three-point correlation function of the comoving curvature
perturbations using Eq.~(\ref{therm-R}) as
\begin{eqnarray}
\left\langle{\mathcal R}_{NL}({\mathbf k}_1){\mathcal R}_{NL}({\mathbf k}_2){\mathcal R}_{NL}({\mathbf k}_3)\right\rangle_{\beta}&\simeq&(2\pi)^{-\frac32}\delta^3({\mathbf k_1}+{\mathbf k_2}+{\mathbf k_3})(2m_{\rm Pl}^2\epsilon)\frac{\partial}{\partial\phi}\left(\frac{H}{\dot{\phi}}\right)\nonumber\\
&\times&\left[\frac{P_{\cal R}(k_1)}{k_1^3}\frac{P_{\cal R}(k_2)}{k_2^3}\left(1+\frac12f_B(k_1)+\frac12f_B(k_2)+f_B(k_1)f_B(k_2)\right)\right.\nonumber\\
&+&\frac{P_{\cal R}(k_2)}{k_2^3}\frac{P_{\cal R}(k_3)}{k_3^3}\left(1+\frac12f_B(k_2)+\frac12f_B(k_3)+f_B(k_2)f_B(k_3)\right)\nonumber\\
&+&\left.\frac{P_{\cal R}(k_3)}{k_3^3}\frac{P_{\cal R}(k_1)}{k_1^3}\left(1+\frac32f_B(k_3)+\frac32f_B(k_1)+3f_B(k_3)f_B(k_1)\right)\right],\nonumber\\
\end{eqnarray}
where $P_{\cal R}(k)$ is defined in Eq.~(\ref{PRk}). The three momenta
form a triangle due to the presence of the delta function. In general
three different triangle configurations are considered to determine
the non-Gaussian effect. The non-linear parameter $f_{NL}$ for these
three momenta configurations are discussed below:
\begin{itemize}
\item {\it Squeezed triangle case} :
For a ``squeezed'' triangle the configuration suggests
  $|{\mathbf k}_1|\approx|{\mathbf k}_2|\approx k\gg|{\mathbf
  k}_3|$. In this configuration the $f_{NL}$ will be
\begin{eqnarray}
f_{NL}^{\rm th}=\frac{5}{6}(\delta-\epsilon)\left(2+2f_B(k_3)\coth\left(\frac{\beta k}{2}\right)\right).
\end{eqnarray}
At low temperature $\beta\rightarrow \infty$ and $f_B(k_3)\rightarrow
0$, yielding the same contribution to the $f_{NL}$ for super-cool
inflation. The minimum value $k_3$ can obtain when the corresponding
wavelength is of Hubble size while crossing the horizon such that
$\lambda_3=\frac{1}{k_3}\sim H^{-1}$ which implies $\beta k_3\sim
0.238$. Hence it yields
\begin{eqnarray}
f_{NL}^{\rm th}&=&\frac{5}{6}(\delta-\epsilon)\times2\left(1+3.72\coth\left(\frac{\beta k}{2}\right)\right)\nonumber \\
&=&f_{NL}\times2\left(1+3.72\coth\left(\frac{\beta
k}{2}\right)\right).
\end{eqnarray}
A lower bound on $\beta k$ can be given from thermal power spectrum
which is given in Eq.~(\ref{lower}) and for this constraint $f_{NL}$
will be maximum yielding $f_{NL}^{\rm th}=64.82f_{NL}\sim0.65$.
\item {\it Equilateral triangle case} :
For a ``equilateral'' triangle we have $|{\mathbf k}_1|=|{\mathbf
  k}_2|=|{\mathbf k}_3|=k$ and in this case the $f_{NL}$ will be
\begin{eqnarray}
f_{NL}^{\rm th}&=&\frac{5}{6}(\delta-\epsilon)\left(3+\frac{5}{4\sinh^2\left(\frac{\beta k}{2}\right)}\right)\nonumber \\
&=&f_{NL}\left(3+\frac{5}{4\sinh^2\left(\frac{\beta k}{2}\right)}\right).
\end{eqnarray}
This implies that for the modes corresponding to our present horizon
$\beta k>0.238$ and the $f_{NL}^{\rm th}=90.85f_{NL}\sim0.9$.

\item {\it Folded triangle case} :
 For ``flattened'' isosceles triangle or  the ``folded''
triangle case we have $|{\mathbf k}_1|=|{\mathbf
  k}_3|=\frac12|{\mathbf k}_2|=k$ and in this case the $f_{NL}$ will
be
\begin{eqnarray}
f_{NL}^{\rm th}&=&\frac{5}{6}(\delta-\epsilon)\left(3+\frac{1}{\sinh^2\left(\frac{\beta k}{2}\right)}\right)\nonumber\\
&=&f_{NL}\left(3+\frac{1}{\sinh^2\left(\frac{\beta k}{2}\right)}\right).
\end{eqnarray}
In this configuration the non-linearity will be $f_{NL}^{\rm
  th}=73.28f_{NL}\sim0.73$ at horizon crossing for the modes
corresponding to our current horizon.
\end{itemize}
\begin{figure}[h]
\centering
\includegraphics[width=0.8\textwidth]{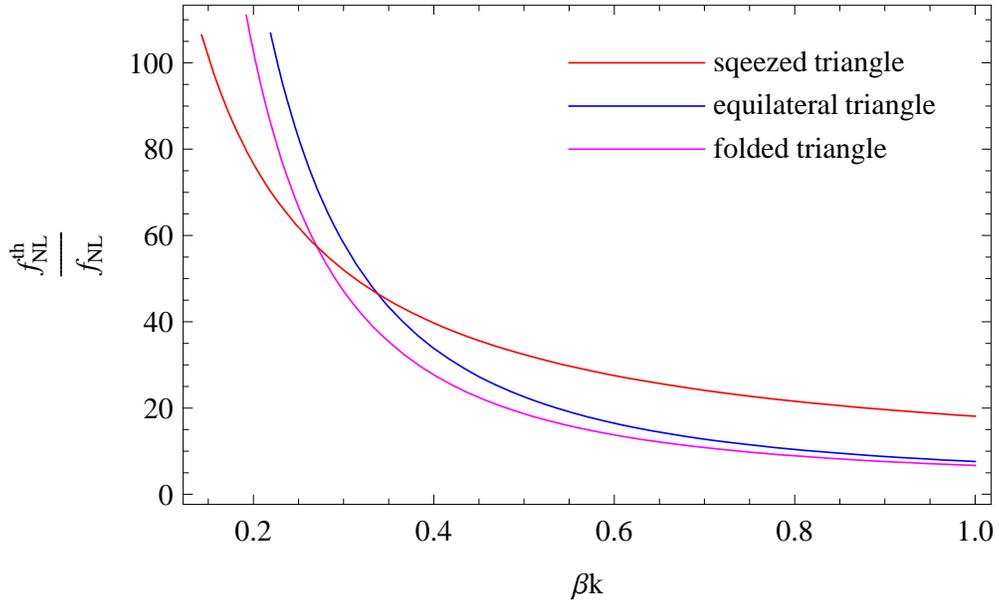}
\caption{$\frac{f_{NL}^{\rm th}}{f_{NL}}$ as a function of ${\beta
  k}$ for different triangle configurations of the three momenta.}
\label{fNL}
\end{figure}
In Fig.~(\ref{fNL}) the thermal enhancement factor $\frac{f_{NL}^{\rm
    th}}{f_{NL}}$ is plotted as a function of ${\beta k}$ for three
different triangle configurations of the three momenta. From the above
discussion it is seen that the maximum contribution for $f_{NL}$ comes
from the ``equilateral'' configuration, though the contribution from
the other two configurations are of the same order. Non-Gaussianity in
all these three cases may be measurable by the 21-cm background
radiation observations \cite{cooray}.

\section{Non-Gaussianity in trispectrum due to  thermal distribution of inflaton}
\label{sec-4}

 In a generic slow-roll single field model the trispectrum
$T(k_1,k_2,k_3,k_4)$ is defined as the Fourier counterpart of the
connected part of four point correlation function of comoving
curvature perturbation \cite{wise}
\begin{eqnarray}
\left\langle{\mathcal R}({\mathbf k}_1){\mathcal R}({\mathbf k}_2){\mathcal R}({\mathbf k}_3){\mathcal R}({\mathbf k}_4)\right\rangle_c=\tau_{NL}\delta^3({\mathbf k_1}+{\mathbf k_2}+{\mathbf k_3}+{\mathbf k_4})T(k_1,k_2,k_3,k_4),
\label{tri}
\end{eqnarray}
where $\tau_{NL}$ is the non-linear parameter for trispectrum and is a
measure of non-Gaussianity. The connected part in the above equation
is defined as
\begin{eqnarray}
\left\langle{\mathcal R}({\mathbf k}_1){\mathcal R}({\mathbf k}_2){\mathcal R}({\mathbf k}_3){\mathcal R}({\mathbf k}_4)\right\rangle_c&=&\left\langle{\mathcal R}({\mathbf k}_1){\mathcal R}({\mathbf k}_2){\mathcal R}({\mathbf k}_3){\mathcal R}({\mathbf k}_4)\right\rangle\nonumber \\
&-&\left(\left\langle{\mathcal R}_{L}({\mathbf k}_1){\mathcal R}_{L}({\mathbf k}_2)\right\rangle\left\langle{\mathcal R}_{L}({\mathbf k}_3){\mathcal R}_{L}({\mathbf k}_4)\right\rangle+2\,\,{\rm perm}\right).
\label{connected}
\end{eqnarray}
The comoving curvature perturbation ${\mathcal R}$ has been expanded
non-linearly upto ${\mathcal O}(\delta\phi_L^2)$. Hence the term
$\left\langle{\mathcal R}_L({\mathbf k}_1){\mathcal R}_L({\mathbf
  k}_2){\mathcal R}_L({\mathbf k}_3){\mathcal R}_{NL}({\mathbf
  k}_4)\right\rangle$ vanishes as it turns out to be a expectation
value of odd number of Gaussian variables $\delta\phi_L$. If one
expands ${\mathcal R}$ upto ${\mathcal O}(\delta\phi_L^3)$ then a term
like $\left\langle{\mathcal R}_L({\mathbf k}_1){\mathcal R}_L({\mathbf
  k}_2){\mathcal R}_L({\mathbf k}_3){\mathcal R}_{NL}({\mathbf
  k}_4)\right\rangle$ will survive and the non-linear parameter
$\tau_{NL}$ will be $\tau_{NL}=\left(\frac65f_{NL}\right)^2$ i.e.
(${\mathcal O}(\epsilon^2)$) \cite{wands}. The trispectrum
$T(k_1,k_2,k_3,k_4)$ in this case will be proportional to product of
three power spectrums and the four momenta form quadrilateral
configuration due to the delta function in Eq.~(\ref{tri}).

But in this model of slow-roll inflation with a radiation era prior to
inflation, the analysis for trispectrum turns out to be quite
different. In presence of a pre-inflationary radiation era the four
point correlation function which contributes to the non-Gaussianity
will be thermal averaged as in the case of power spectrum and
bispectrum. It is worth to point out that due to thermal averaging the
four point function is not just the square of the two-point function
as that would have been the case at zero temperature. So in this case,
by connected part of the four-point function defined in
Eq.~(\ref{connected}) we will simply mean the excess of the thermal
averaged of four-point function than the square of its two-point
Gaussian part and will now define the non-linear parameter $\tau_{NL}$
in the following way
\begin{eqnarray}
\left\langle{\mathcal R}_{L}({\mathbf k}_1){\mathcal R}_{L}({\mathbf k}_2){\mathcal R}_{L}({\mathbf k}_3){\mathcal R}_{L}({\mathbf k}_4)\right\rangle_c&\equiv&\left\langle{\mathcal R}_{L}({\mathbf k}_1){\mathcal R}_{L}({\mathbf k}_2){\mathcal R}_{L}({\mathbf k}_3){\mathcal R}_{L}({\mathbf k}_4)\right\rangle_\beta \nonumber\\
&-&\left(\left\langle{\mathcal R}_{L}({\mathbf k}_1){\mathcal R}_{L}({\mathbf k}_2)\right\rangle_\beta\left\langle{\mathcal R}_{L}({\mathbf k}_3){\mathcal R}_{L}({\mathbf k}_4)\right\rangle_\beta+2\,\,{\rm perm.}\right)\nonumber\\
&=&\tau_{NL}\left[\frac{P_{\mathcal R}(k_1)}{k_1^3}\frac{P_{\mathcal R}(k_2)}{k_2^3}\delta^3({\mathbf k_1}+{\mathbf k_3})\delta^3({\mathbf k_2}+{\mathbf k_4})\right.\nonumber\\
&&\,\,\,\,\,\,\,\,\,\,\left.+2\,\,{\rm perm.}\right].
\label{tau}
\end{eqnarray}
Hence in this case $\tau_{NL}$ will not depend upon the slow-roll
parameters. The thermal average of the four-point correlation function
of inflaton fluctuation has been calculated in the last section in
Eq.~(\ref{fpc}). Using this equation the thermal average of the
four-point correlation of curvature perturbation can be derived as
\begin{eqnarray}
&&
\left\langle{\mathcal R}_{L}({\mathbf k}_1){\mathcal R}_{L}({\mathbf k}_2){\mathcal R}_{L}({\mathbf k}_3){\mathcal R}_{L}({\mathbf k}_4)\right\rangle_\beta=\frac{P_{\mathcal R}(k_1)}{k_1^3}\frac{P_{\mathcal R}(k_2)}{k_2^3}\left[\delta^3({\mathbf k_1}+{\mathbf k_4})\delta^3({\mathbf k_2}+{\mathbf k_3})\right.\nonumber \\
&&+\delta^3({\mathbf k_1}+{\mathbf k_3})\delta^3({\mathbf k_2}+{\mathbf k_4})
\left.\left\lbrace1+f_B(k_1)+f_B(k_2)+2f_B(k_1)f_B(k_2)\right\rbrace\right]+\frac{P_{\mathcal R}(k_1)}{k_1^3}\frac{P_{\mathcal R}(k_3)}{k_3^3}\nonumber\\
&&\times\left[\delta^3({\mathbf k_1}+{\mathbf k_2})\delta^3({\mathbf k_3}+{\mathbf k_4})\left\lbrace1+2f_B(k_1)+2f_B(k_3)+4f_B(k_1)f_B(k_3)\right\rbrace\right],
\end{eqnarray}
and the thermal average of two-point function can be given in terms of
the power spectrum as
\begin{eqnarray}
\left\langle{\mathcal R}_{L}({\mathbf k}_1){\mathcal R}_{L}({\mathbf k}_2)\right\rangle_\beta=\frac{P_{\mathcal R}(k_1)}{k_1^3}\left(1+2f_B(k_1)\right)\delta^3({\mathbf k_1}+{\mathbf k_2}).
\end{eqnarray}
Hence the connected part will be
\begin{eqnarray}
&&\left\langle{\mathcal R}_{L}({\mathbf k}_1){\mathcal R}_{L}({\mathbf k}_2){\mathcal R}_{L}({\mathbf k}_3){\mathcal R}_{L}({\mathbf k}_4)\right\rangle_c=-\frac{P_{\mathcal R}(k_1)}{k_1^3}\frac{P_{\mathcal R}(k_2)}{k_2^3}\left[\delta^3({\mathbf k_1}+{\mathbf k_3})\delta^3({\mathbf k_2}+{\mathbf k_4})\left\lbrace f_B(k_1)+\right.\right.\nonumber\\
&&f_B(k_2)\left.\left.+2f_B(k_1)f_B(k_2)\right\rbrace+2\delta^3({\mathbf k_1}+{\mathbf k_4})\delta^3({\mathbf k_2}+{\mathbf k_3})\left\lbrace f_B(k_1)+f_B(k_2)+f_B(k_1)f_B(k_2)\right\rbrace\right].\nonumber\\
\label{tau-th}
\end{eqnarray}
The four momenta in this case will not form a quadrilateral as in
other trispectrum cases. But due to the presence of two delta
functions on the R.H.S. of the above equation the non-linear parameter
$\tau_{NL}$ can be calculated in the following two cases :
\begin{enumerate}
\item ${\mathbf k_1}=-{\mathbf k_3}$, ${\mathbf k_2}=-{\mathbf k_4}$
  and $k_i=k\,\,(i=1,2,3,4)$ :
\begin{eqnarray}
\tau_{NL}^{\rm th}=-\frac{1}{\cosh(\beta k)-1}.
\end{eqnarray}
The maximum observable value of $\left|\tau_{NL}^{\rm th}\right|$ can
be obtained using the constraint on the comoving temperature as $\beta
k>0.238$. Hence for $\beta k\sim0.238$ one finds that $\tau_{NL}^{\rm
  th}\sim-35.14$.
\item ${\mathbf k_1}=-{\mathbf k_4}$, ${\mathbf k_2}=-{\mathbf k_3}$
  and $k_i=k\,\,(i=1,2,3,4)$ :
\begin{eqnarray}
\tau_{NL}^{\rm th}=-2\frac{1-2e^{\beta k}}{\left(e^{\beta k}-1\right)^2}.
\end{eqnarray}
The maximum value of $\tau_{NL}^{\rm th}$ for this case will be
$\tau_{NL}^{\rm th}\sim-42.58$.
\end{enumerate}

\begin{figure}[h]
\centering
\includegraphics[width=0.8\textwidth]{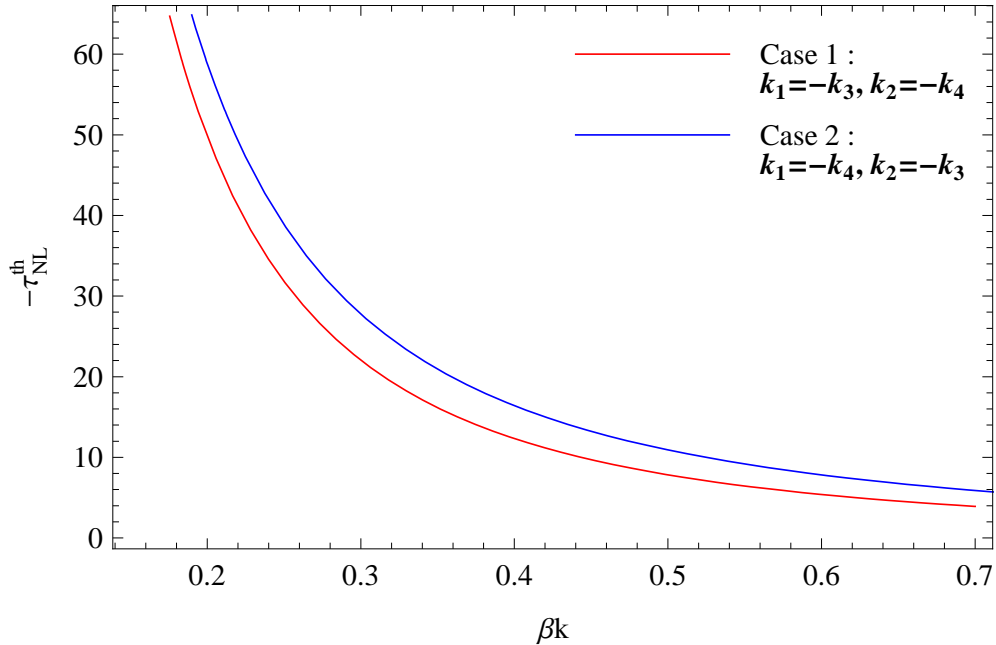}
\caption{Plot of $\tau_{NL}^{\rm th}$ for two different momenta
  configurations as a function of $\beta k$}
\label{tau-fig}
\end{figure}
In Fig.~(\ref{tau-fig}) we have plotted $\tau_{NL}^{\rm th}$ as a
function of $\beta k$. We find that the maximum value of
non-Gaussianity comes from the configuration when ${\mathbf
  k_1}=-{\mathbf k_4}$, ${\mathbf k_2}=-{\mathbf k_3}$ and
$k_i=k\,\,(i=1,2,3,4)$ which is $\sim-42$. We do not compare this
contribution to non-Gaussianity due to thermal initial states with the
zero temperate case as there is no contribution from the later at this
order and the leading order $\tau_{NL}$ in zero temperature is
${\mathcal O}(\epsilon^2)$.
\section{Conclusion} 

We studied the effect of a decoupled thermal spectrum of inflatons
(which exist in the scenario where the inflation is preceded by a
prior thermal era) on the non-Gaussianity of the primordial
perturbation. We found that thermal inflatons can enhance the
bispectrum non-Gaussianity parameter $f_{NL}$ by a factors of
$(65-90)$ depending upon the momentum configuration. The zero
temperature non-Gaussianity parameter $f_{NL}$ in single field
inflation models is proportional to the slow roll parameters and is
expected to be of order $\sim 10^{-2}$. Therefore the observed value
of $f_{NL}$ in thermal history models will be of $\sim 1$. This is too
small to be measured by WMAP or even the forthcoming PLANCK
experiment. Measurements of anisotropies in the Hydrogen 21-cm
radiation background can detect non-Gaussianities as low as $f_{NL}
\sim 0.1$ \cite{cooray}, and this may be the ideal experiment in which
non-Gaussianities with a thermal origin can be observed. The 21-cm
observations may also be able to measure the non-Gaussianity in the
trispectrum $\tau_{NL}\sim{\mathcal O}(10)$ and for which the
prediction from thermal history inflation scenarios is $0>\tau_{NL}>
-43$. We conclude that a signature of thermal inflaton background at
the time of inflation is a large trispectrum non-Gaussianity compared
to the bispectrum non-Gaussianity.
 
\appendix
\section{Non-Gaussianity in a single-field slow-roll inflation model}
\label{app-1}

In a single field slow-roll model Non-Gaussianity appears generically
once the inflaton field has self-interactions such as
$V(\phi)\sim\phi^3$ or $V(\phi)\sim\phi^4$. But the Non-Gaussianity
due to these self-interactions is very small where the bispectrum
Non-Gaussian parameter $f_{\rm NL}\sim{\mathcal
  O}(\epsilon^2)$. Larger contribution to Non-Gaussianity in such
models come from non-linear curvature perturbations and in this cases
$f_{\rm NL}\sim{\mathcal O}(\epsilon,\eta)$ \cite{komatsu}. Here we
will briefly discuss how non-linearity in comoving curvature
perturbation gives rise to Non-Gaussianity of the order of slow-roll
parameters $\epsilon$ and $\eta$.

The quantum fluctuations in inflaton field generates fluctuation in
the metric which is coupled to it through Einstein's equation. The
perturbed FRW metric has the form (considering only scalar
perturbations)
\begin{eqnarray}
\tilde{g}_{\mu\nu}=a^2(\eta)\left(
\begin{array}{ccc}
1+2A & & 0 \\
0 & & -(1-2\psi)\delta_{ij}
\end{array}
\right),
\end{eqnarray}
where the quantity $\psi$ is known as the spatial curvature
perturbation, $\eta$ is the conformal time and $a(\eta)$ is the cosmic
scale factor. The gauge invariant quantity formed out of this spatial
curvature perturbation is known as the comoving curvature perturbation
and defined as
\begin{eqnarray}
{\mathcal R}(t,{\mathbf x})=\psi(t,{\mathbf x})+\frac{H}{\dot{\phi}}\delta\phi(t,{\mathbf x}).
\end{eqnarray}
These perturbations are conserved on super-horizon scales throughout
the evolution. Hence during inflation in the spatially flat gauge this
reduces to
\begin{eqnarray}
{\mathcal R}(t,{\mathbf x})=\frac{H}{\dot{\phi}}\delta\phi(t,{\mathbf x}),
\label{R}
\end{eqnarray}
and after inflation when $\delta\phi\sim 0$ this represents the
gravitational potential on comoving hypersurfaces
\begin{eqnarray}
{\mathcal R}(t,{\mathbf x})=\psi(t,{\mathbf x}).
\end{eqnarray}
The CMB anisotropy spectrum is determined by the power spectrum of
this comoving curvature perturbation which is related to the power
spectrum of scalar perturbation as
\begin{eqnarray}
{\cal P}_{\cal R}(k)=\frac{k^3}{2\pi^2}\langle{\mathcal R}(k){\mathcal R}(k)\rangle=\frac{1}{2m_{\rm Pl}^2\epsilon}P_{\delta\phi}(k),
\label{curvature}
\end{eqnarray}
where the slow-roll parameter $\epsilon\equiv 4\pi
G\frac{\dot{\phi}^2}{H^2}$, $m_{\rm Pl}\equiv\frac{1}{\sqrt{8\pi G}}$
is the reduced Planck mass and $G$ being the Newton's Gravitational
constant.

Presuming that the inflaton fluctuations $\delta\phi$ are initially
Gaussian, the comoving curvature perturbations $\mathcal {R}$ given in
Eq.~(\ref{R}) also obeys Gaussian statistics in the linear order
\begin{eqnarray}
{\mathcal R}_L(t,{\mathbf x})=\frac{H}{\dot{\phi}}\delta\phi_L(t,{\mathbf x}),
\label{RL}
\end{eqnarray}
where ${\mathcal R}_L(t,{\mathbf x})$ and can be expanded in Fourier
space as
\begin{eqnarray}
{\mathcal R}_L(t,{\mathbf x})=\int\frac{d^3{\mathbf k}}{\left(2\pi\right)^{\frac32}}e^{i{\mathbf k}\cdot{\mathbf x}}{\mathcal R}_L(t,{\mathbf k}).
\end{eqnarray}
Being constant in time outside the horizon these comoving curvature
fluctuations after entering the horizon in later times produces
curvature perturbations which are Gaussian in nature.

In the non-linear limit one observes that \cite{komatsu}
$\frac{H}{\dot{\phi}}\equiv-\frac{1}{m_{\rm Pl}^2}\frac{V\left(\phi\right)}{V^\prime\left(\phi\right)}$
is a function of $\phi$ and hence
\begin{eqnarray}
{\mathcal R}_{NL}(t,{\mathbf x})=\frac{H}{\dot{\phi}}\delta\phi_L(t,{\mathbf x})+\frac12\frac{\partial}{\partial\phi}\left(\frac{H}{\dot{\phi}}\right)\delta\phi_L^2(t,{\mathbf x})+{\mathcal O}(\delta\phi_L^3). 
\label{RNL}
\end{eqnarray}
Therefore in the Fourier space one gets
\begin{eqnarray}
{\mathcal R}_{NL}(t,{\mathbf k})=\frac{H}{\dot{\phi}}\delta\phi_L(t,{\mathbf k})+\frac12\frac{\partial}{\partial\phi}\left(\frac{H}{\dot{\phi}}\right)\int\frac{d^3{\mathbf p}}{\left(2\pi\right)^{\frac32}}\delta\phi_L(t,{\mathbf p})\delta\phi_L(t,{\mathbf k}-{\mathbf p}),
\label{Rk}
\end{eqnarray}
where $\delta\phi_L(t,{\mathbf k})$ has the same form as given in
Eq.~(\ref{phik}).

The non-linear parameter $f_{NL}$ for bispectrum $B(k_1,k_2,k_3)$ or
the three-point correlation function of the comoving curvature
perturbation is defined as \cite{wise}
\begin{eqnarray}
\left\langle{\mathcal R}({\mathbf k}_1){\mathcal R}({\mathbf k}_2){\mathcal R}({\mathbf k}_3)\right\rangle&=&(2\pi)^{-\frac32}\delta^3({\mathbf k_1}+{\mathbf k_2}+{\mathbf k_3})\frac65 f_{NL}\left(\frac{P_{\cal R}(k_1)}{k_1^3}\frac{P_{\cal R}(k_2)}{k_2^3}+2\,\,{\rm perms.}\right)\nonumber\\
&=&(2\pi)^{-\frac32}\delta^3({\mathbf k_1}+{\mathbf k_2}+{\mathbf k_3})\frac65 f_{NL}B(k_1,k_2,k_3),
\label{bispec}
\end{eqnarray}
where we have used the definition
\begin{eqnarray}
P_{\cal R}(k)=(2\pi^2){\cal P}_{\cal R}(k),
\label{PRk}
\end{eqnarray}
and the normalization $(2\pi)^{-\frac32}$ has been chosen accordingly.
Hence using Eq.~(\ref{Rk}) one can compute the bispectrum in this case
as follows
\begin{eqnarray}
\left\langle{\mathcal R}_{NL}({\mathbf k}_1){\mathcal R}_{NL}({\mathbf k}_2){\mathcal R}_{NL}({\mathbf k}_3)\right\rangle\simeq&&(2\pi)^{-\frac32}\delta^3({\mathbf k_1}+{\mathbf k_2}+{\mathbf k_3})(2m_{\rm Pl}^2\epsilon)\frac{\partial}{\partial\phi}\left(\frac{H}{\dot{\phi}}\right)\nonumber\\
&&\times\left(\frac{P_{\cal R}(k_1)}{k_1^3}\frac{P_{\cal R}(k_2)}{k_2^3}+2\,\,{\rm perms.}\right).
\end{eqnarray}
Comparing the above two equations the non-linearity parameters in this
case will be
\begin{eqnarray}
f_{NL}&=&\frac53 m_{\rm Pl}^2\epsilon\frac{\partial}{\partial\phi}\left(\frac{H}{\dot{\phi}}\right)\nonumber\\
&=&-\frac{5\epsilon}{3}\frac{\partial}{\partial\phi}\left(\frac{V(\phi)}{V^\prime(\phi)}\right).
\label{fnl}
\end{eqnarray}
The non-linear parameter $f_{NL}$ given in Eq.~(\ref{fnl}) can be
fully expressed in terms of the slow-roll parameters
$\epsilon\equiv\frac{1}{16\pi G}\left(\frac{V^{\prime}}{V}\right)^2$,
$\eta\equiv\frac{1}{8\pi G}\left(\frac{V^{\prime\prime}}{V}\right)$
and $ \delta\equiv \eta-\epsilon$ as
\begin{eqnarray}
f_{NL}=\frac56(\delta-\epsilon).
\label{fnl-slow}
\end{eqnarray}
The form of the $f_{NL}$ derived above is same as in
\cite{wands}. Here $f_{NL}$ being proportional to the slow-roll
parameters is too small to be detected by the ongoing experiments.

The non-linearity parameter can also be expressed in terms of the
potential $V(\phi)$ using Eq.~(\ref{fnl}) as
\begin{eqnarray}
f_{NL}=-\frac56m_{\rm Pl}^2\left(\frac{V^\prime}{V}\right)\frac{\partial}{\partial\phi}\left(\frac{V(\phi)}{V^\prime(\phi)}\right).
\end{eqnarray}
This equation is useful in the case where the form of the potential is
known. Such as, for a power-law potential where $V(\phi)\sim\phi^n$
the non-linearity parameter will be
\begin{eqnarray}
f_{NL}=-\left(\frac56\right)n\frac{m_{\rm Pl}^2}{\phi^2},
\end{eqnarray}
which is same as the one given in \cite{komatsu}.

\end{document}